# Masses of the Sub-Nuclear Particles

Theodore M. Lach

**The masses of the quarks and leptons are for the most part a mystery to particle physicists. Currently there seems to be no correlation between the masses of the elementary particles. This paper is an attempt to formulate a theory that would begin to explain the relationship between the sub-nuclear particle masses.**

A key premise of the Standard Model is that there are 3 and only 3 generations of quarks and leptons. The currently accepted masses of these elementary particles are: (*1*)

### Table 1.

| Generation # 1 | Generation # 2 | Generation # 3 |
|---|---|---|
| Up quark = 3 MeV, (1-5) | Charmed quark = 1.25±0.1 GeV | Top quark = 175 ±5 GeV |
| Down quark = 6 MeV (3-9 MeV) | Strange quark = 100 MeV (75-170 MeV) | Bottom quark = 4.2 GeV ± 0.2 GeV |
| Electron = 0.5109989 MeV | Muon = 105.6583 MeV | Tau = 1777.1 MeV |

The neutrino masses in the standard model are still a hotly debated issue, yet most current references put the upper bound of the three neutrinos at:

### Table 2.

| | |
|---|---|
| Electron neutrino = 10 ev | (5 – 17 ev) |
| Muon neutrino = 270 KeV | (200 – 270 KeV) |
| Tau neutrino = 31 MeV | (20 – 35 MeV) |

In an attempt to determine why the quark masses predicted in the CBM (2), Checker Board Model of the nucleus, do not fit into the standard model, I will deviate from the S.M. and assume that my quark masses in table 3 are correct and that there are more than 3 generations.

**Table 3.**

| "up" Quark = 237.31 MeV |
| --- |
| "dn" Quark = 42.39 MeV |

I will try to fit these new masses in table 3. into a self-consistent theory of the masses of the sub nuclear particles. In the rest of this paper I will use the terms "up" and "dn" to refer to my predictions of the mass of the quarks in the proton and neutron, and will use the terms "u" and "d" to refer to what up until now the S.M. has called up and down. This paper will intend to prove that the u and d quarks do exist but they are not the same as the up and dn quarks. In this paper I will use the term family of up quarks or family of dn quarks to refer to all 2/3 or – 1/3 charged quarks in this model. I will also use the term electron family and neutrino family for the same purpose.

The first step in this process is to assume that a NEW generation of quarks exists between the electron generation and the muon generation. I came to this conclusion slowly over many years of consideration. This resulted in a void for the lepton particles that go along with the up and dn quarks. I thought a name for these "new" leptons should be "nu" (pronounced "new", and the Greek letter for n or "ν"). Consequently, the new particle in the electron family could be called the nuon. To fit into this pattern or structure, I decided the nuon must have a mass of about 7.57 MeV. Figure 1 shows a curve fit of the 4 lepton masses including the newly hypothesized 7.57 MeV nuon.

Recent discovery of neutrino oscillations have led to the hypothesis that at least one new "sterile" neutrino must exist. (*3-6*) Some theorists put the mass of this new neutrino between the mass of the electron and the mass of the muon neutrino. In a recent special issue of Fermi News in an article by Mike Perricone, he states that there is a quip among neutrino circles, no one knows who said it first, but it says: "Sterile

neutrinos are like cockroaches. Once you get one in your theory, there's no stopping them."(*7*)

Interestingly, back in 1985, John Simpson, a Canadian Physicist at the University of Guelph, thought he had discovered a new neutrino with a mass of about 17 KeV. (*8,9*) In 1991 and 1992 there was a great debate about whether to accept this data. At the end of this debate, the proponents of the standard model apparently convinced everyone that this "Neutrino from Hell" would not fit into the standard model and it needed to be abolished. Stuart Freedman, a very well respected neutrino physicist, is apparently quoted as saying that he is still impressed that more than six different kinds of experiments have shown an effect, and all at the 17 KeV mark. "It is very surprising that different experiments get the same value for these things. It is hard to believe that it's a bunch of random systematic errors."(*10*)

Therefore, my justification for adding this new generation is based upon: 1. The up and dn quarks predicted by CBM are too heavy to be the u and d quarks, 2. The suspicion that there is a new neutrino (sterile or not) between the electron and muon neutrino to help explain neutrino oscillations, and lastly 3. The fit of the leptons along a straight line in Figure 1. (It is interesting to note that slope of this line (ln (0.511 / 1777.1) = -3e) represents a slope of "e" to better than 1 part in 10,000.)

Next, I accept the S.M. masses of the charmed and bottom quarks. Since these particles are so heavy, their relativistic effects within the $J/\psi$ meson and the upsilon ϒ meson are small and therefore we can use non-relativistic Q.M. to determine the intrinsic mass of these two quarks.(11) Seeing how the four leptons fit on a smooth curve in Figure 1, a similar technique was used for the dn like quarks. Using the mass of dn at 42.39 MeV and the mass of the bottom quark at $4238 \pm 006$ MeV based upon the work of Voloshin (12), it seemed that the strange quark should be 424 MeV and the d quark should be 4.2 MeV. A mass of 4.2 MeV for the d quark is close to the standard model's prediction of 3-9 MeV, since the uncertainty of this estimate is very model dependent. See Figure 2 for a plot of the dn family quarks. It seemed interesting that when the curves for the electron family and dn quark family were extended, they seemed to converge at about 413-424 GeV. I believe most theorists expect a convergence like this to occur, resulting in the grand unification at a super massive long sought after Higgs Boson.

The mass differences of the $K^-$ and $K^0$, and the mass differences of the $\pi^-$ and $\pi^0$ mesons suggest that there is a "dn-like quark" which must be 4 MeV more massive than an "up-like quark". My model would suggest that the dn quark (42.39 MeV) is 4 MeV heavier than the u quark. (38 MeV). Using the 424 GeV point as a pivot point, the mass of the up quark, (237.31 MeV) and 1500 MeV for the mass of the Charmed quark based upon Alvarex (*13*), one can then draw a straight line through these four points. (see Figure 2) This value of the mass of the u quark is significantly outside the range of the S.M. theoretical estimate of 1-5 MeV.

The reader will note that a massive quark generation (generation #1) was inserted between the 424 GeV Higgs mass and the Tau generation (generation #2). Quoting the quip from before, once you let one of those cockroaches in, more will be sure to follow. Figure 2 suggests that there is a 67 GeV massive up quark (Mup), and a massive dn quark (Mdn) of 42.4 GeV and an associated electron family member of 27 GeV. In support of this new massive generation, a paper was published in 1984 by over 130 physicists as part of the UA1 collaboration at CERN. This paper established experimental data for the existence of a quark with mass between 30 and 50 GeV, with the most likely value at about 40 GeV.(*14*)  This finding was later rejected, but as one can see from the proceedings of the first Aspen Conference on High Energy Physics in

1985, the report was much discussed and appeared at that time to have significant supporting evidence.(*15*)

One additional interesting feature of this new model is that the curves of best fit though the data points are equally spaced. Note the exponent of the curve fitting trend lines differ from one another by an approximate amount 0.435 – 0.425.

I now refer the reader to figure 3. Here I have added the best estimates of the upper bounds of the neutrino masses to the curves in figure 2. To first order a straight line through these upper bound masses seem to extrapolate to the same point (Higgs Mass), yet the fit is off by about 75 GeV at generation # zero, a small difference on the scale of this plot. This degree of deviation is not too surprising since the upper bounds of the estimates of the neutrino masses change quite often. I suggest the reader notice how nicely an opening for a 1.7 KeV neutrino (a John Simpson like neutrino) would fit into this model. Although this 1.7 KeV is an order of magnitude lower than what Simpson found, it must be understood that experimental refinements in neutrino mass determinations usually end up with the refined numbers being lower.

I need to say something about the top quark. This model predicts a mass of the top quark of 10 GeV. This is significantly at odds with the particle / resonance discovered at Fermi Lab in 1995 that determined the mass of the top quark at 175 GeV. My model does not suggest what the 175 GeV particle might be (perhaps a combination of massive quarks from generation 1, perhaps two Mup and one Mdn), and since the half-life of this excited state is only 0.4 yoctoseconds (0.4 X $10^{-24}$ seconds) (*16*) it is hard to make detailed studies of this particle. I suspect that the reason the top quark in my model has not been found is that its mass is almost identical with (and therefore hidden by) the Y (S4) upsilon meson which decays into two bottom quarks (sometimes referred to as naked bottom). This may be why the ϒ (S4) meson resonance is broad, unlike the other upsilon resonances and also is not tied to the other resonances by photon decay transitions. (*17*)

This model predicts a mass of the strange quark significantly heavier and outside the range of the currently accepted value (75-170 MeV) based upon the standard model. Michael Shaevitz, a member of the NuTeV team, is quoted as saying, "what we have found is that the strange content is much smaller than expected [based upon the standard model], about one-half the amount of the up or down quark sea." (*18*) The hope is that this new predicted mass of the strange quark would help explain this finding. Also regarding the sea of quarks in the proton, "Michael Leitch (of the Los Alamos National Laboratory) and his colleagues on another Fermilab experiment, nicknamed NuSea, [in 1998] uncovered an even more startling inconsistency [in the standard model]. The number of antiup quarks in the proton sea is not the same as the number of antidowns." (*18*) Perhaps the predicted masses of the "up" and "dn" quarks will resolve this problem also.

Is there any justification for a Higgs Mass at 424 GeV? In 1996, according to William Carithers the co-spokesperson of the CDF collaboration at Fermilab, CDF had "observed an unexpectedly large number of 'hard,' or violent collisions between quarks, …This is just the sort of effect you would see if the quarks were not fundamental particles but had some sort of internal structure." According to Cuido Altarelli of CERN, "If quark substructure was true, then its relevance [the CDF findings] would be very, very large". (*19*) What was found in this experiment were two jets, one with an energy of 424 GeV. Could this be the first evidence of the Higgs Mass this theory predicts? Is it just coincidence that 424 GeV is almost exactly $10^4$ time larger than the predicted mass of the dn quark and $10^2$ times larger than the most precise estimate of the bottom quark? It is interesting that the initial slope of the electron family line (ln (27,000/424,000)) is approximately "e" the natural logarithm.

It is interesting that the three curves for the sub-nuclear particles (with the exception of the neutrinos) fit so neatly onto equally separated curves on a semi log plot. What this implies is that there is a simple geometric mean relationship that ties all of these sub-nuclear particles together. Example: 42.39 is the geometric mean of 237.31 and 7.57 MeV (That is how the value of 7.57 was selected). 7.74 MeV would have been selected if we wanted to exactly fall on the exponential "e" slope line between the Tau and the electron. Also within a particle family (say dn quarks) 42.39 is the approximate geometric mean of 424 and 4.24 MeV.

Therefore all the masses of the sub-nuclear particles fit into one equation:

$$M_{(F, G)} = H \, e^{-k G} \quad (1)$$

where:

$k_1$ (F = 1 for the up quarks) = 1.8693

$k_2$ (F = 2 for the dn quarks) = 2.304

$k_3$ (F = 3 for the electron family) = 2.7287

G = Generation number for the particle: for Higgs G = zero, for Electron G = 5

H is the mass of the Higgs boson.

F is the family number of the particle type.

One concern is that the muon does not fall on the line as well as it should. It is about 12 MeV below the line, but perhaps this is due to corrections needed to be made to experimental data, based upon the existence of a new $7.6 \pm 0.2$ MeV lepton.

This equation can also be rewritten as: log (m/H) = constant for a given family. As Frank Close points out in the book "The New Physics", "Renormalization requires a dimensional parameter to set the scale of the logarithm. This is called the renormalization scale [in my theory that value is H]. It is arbitrary in principle, but in fact, for any given calculation, some renormalization scales are more convenient than others because of the presence of logs of m/H. If all the masses and momenta in a process are of the same order of magnitude, it pays to choose [H] in the range to minimize the effects of the logs and make the perturbation theory better behave. It is this logarithmic dependence on the renormalization scale which is responsible for the renormalisation group dependence of parameters on the distance or momentum scale, first discussed by Gell-Mann and Low." (20)  Therefore, from a dimensional analysis it seems like a log (m/H) function of mass, has some basis for being considered part of this theory. Whether this model suggests a $6^{th}$ generation (and more cockroaches) perhaps related to super cold temperatures, is difficult to speculate.

This theory also does a better job explaining the low mass mesons. The current standard model has established that the Upsilon meson (Υ (S1) ) is a meson of two bottom quarks and it has a mass of $9460.37 \pm 0.21$ MeV. Also the standard model has established that the meson combined of two charmed quarks is the eta sub c ($\eta_c$) and has a mass of 2980 MeV. Using the accepted masses of the Charmed and Bottom quarks in table 1 we get:

**Table 4.**

| Meson | Quarks | Mass of meson MeV | Mass of 2 Quarks MeV | Delta Mass MeV | Delta mass as % of meson Mass |
|---|---|---|---|---|---|
| Υ (S1) | b $\underline{b}$ | 9460.37 | 8400 | 1060 MeV | 11 % |
| $\eta_c$ | c $\underline{c}$ | 2980 | 2500 | 480 | 16 % |

(Note: I am using the nomenclature b̄ to indicate the anti particle of the b quark.)

Therefore the standard model accepts the mass of the composite meson to be 11% to 16% higher than the rest mass of its constituent quarks, at least for these two examples. How does this process work for the lighter meson?

Table 5.

| Meson | Quarks | Mass of Meson MeV | Mass of two Quarks MeV | Delta Mass MeV | Delta Mass as % of meson Mass |
|---|---|---|---|---|---|
| Eta primed | s s̄ | 957.77 | 200 | 758 | 79 % |
| Eta | u ū | 547.30 | 6 | 541 | 99 % |

As we can see the % delta mass is not anywhere near 11% to 16%. To be fair to the standard model it believes the eta meson is a mixture of states $(u\underline{u} - d\underline{d})/2^{1/2}$. Since the mass of both the u and d in the standard model is very low, the above calculation is approximately correct.

Look how the data looks if I substitute the masses in this theory:

Table 6.

| Meson | Quarks | Mass of Meson MeV | Mass of two Quarks MeV | Delta Mass MeV | Delta Mass as % of meson Mass |
|---|---|---|---|---|---|
| Eta primed | s s̄ | 957.77 | 848 | 110 | 11.5 % |
| Eta | up ūp | 547.30 | 474.6 | 72.7 | 13 % |

Notice with the new mass of up and s from this theory, the numbers in table 6, appear to have the same % delta mass excess as the $\eta_c$ and $\Upsilon$ (S1) heavier mesons. This adds additional support for a strange quark of 424 MeV, which is significantly different than what the S.M predicts. Finally let me show how the $\pi^-$ and $\pi^0$ fit into this model.

**Table 7.**

| Meson | Quarks | Mass of Meson MeV | Mass of two Quarks MeV | Delta Mass MeV | Delta Mass as % of meson Mass |
|---|---|---|---|---|---|
| $\pi^-$ | dn u̲ | 139.5699 | 42+38 = 80 | 60 | 43 % |
| $\pi^0$ | u u̲ | 134.9764 | 38 X 2 = 76 | 59 | 44 % |

Note the delta mass % are not in the 11% to 16% range anymore, but these are very light mesons, and that may account for this change. I would like to point out how this theory has a rest mass of the dn - u̲ meson (80 MeV) that is 4 MeV more massive than the two u quarks (76 MeV) of the $\pi^0$, as it should be. I believe these few examples help add additional credibility to this new model.

**Table 8.**

| Meson | Quarks | Mass of Meson MeV | Mass of two Quarks MeV | Delta Mass MeV | Delta Mass as % of meson Mass |
|---|---|---|---|---|---|
| $\rho^+$ | u s̲ | 770 | 237+424=661 | 109 | 14 % |
| $B^0$ | s b̲ | 5379 MeV | 424 + 4240 | 715 | 13 % |

On a more speculative note, table 8 shows how the rho plus and B zero mesons fit into this pattern assuming they are composed of the strange –1/3 quark instead of the d (– 1/3) quark.

If this model is correct, we have not yet found the lightest meson, the d d̲ meson. It should have a mass of about (4.2 X 2 + 50%) or about 12 MeV. Perhaps it should be called $\eta_d$. Perhaps $\eta'$ (primed) should be renamed $\eta_s$ and $\eta$ should be renamed $\eta_{up}$. $\pi^0$ could be renamed $\eta_u$ but that would be more confusing than helpful.

Conclusion:

This paper attempts to present a self-consistent theory of the masses of sub nuclear particles based upon the assumption that the masses of the up and dn quarks as determined in the CBM of the nucleus are correct. However, all this effort is academic unless a 7.6 $\pm$ 0.2 MeV nuon lepton can be found. This may be difficult since a number of low energy nuclear phenomenon exist in this energy range, for example the typical binding energy per nucleon to name just one. The validity of this theory may also be strengthened by the 7$^{th}$ rediscovery of the John Simpson neutrino especially if turns out to have a mass closer to 1.7 KeV than 17 KeV.

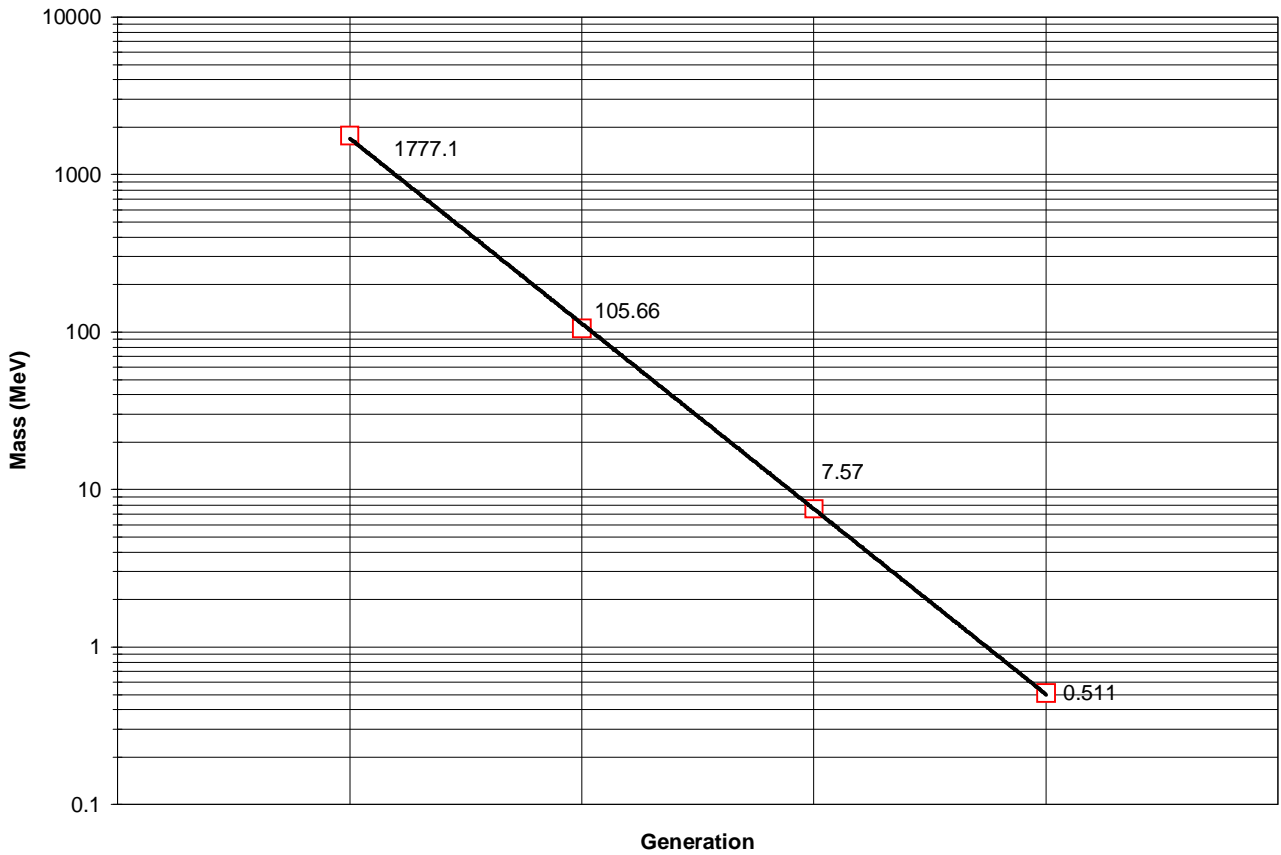

**Mass MeV vs. Generation Number**
**Figure 1**

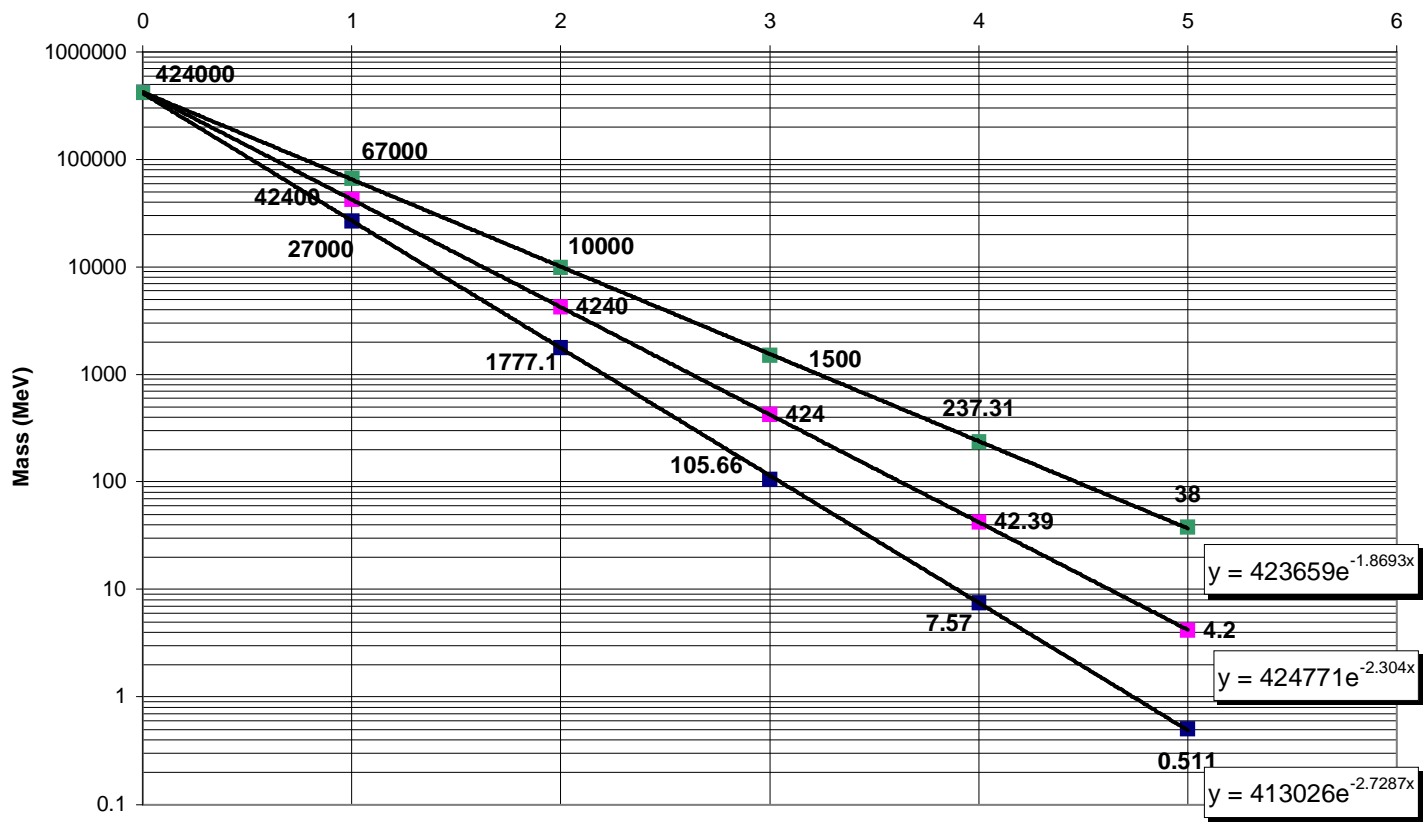

**Masses of Sub-Nuclear Particles**
Figure 2.

# Masses of Sub-Nuclear Particles
## Figure 3.

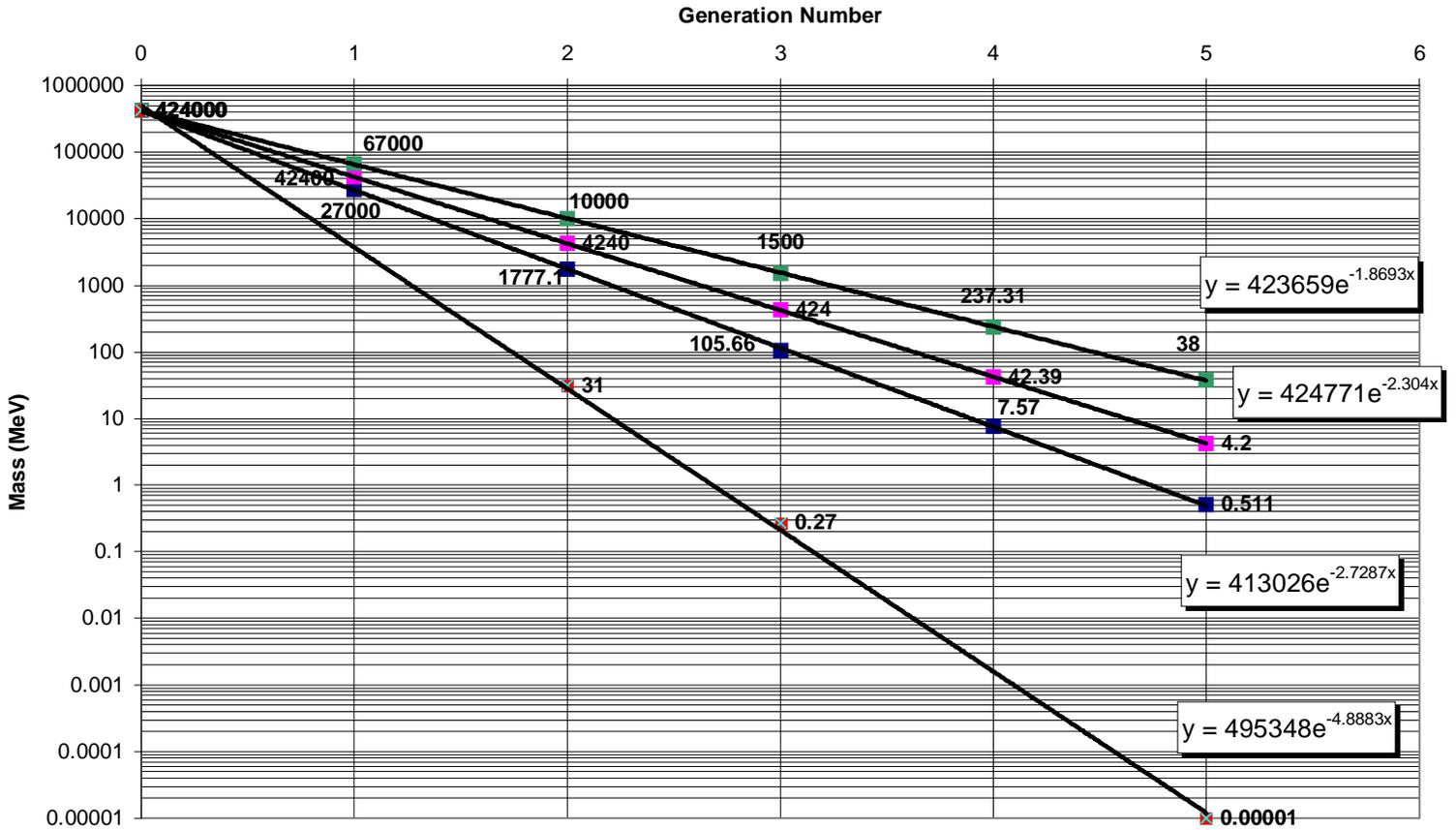